\documentclass[
  aps,
  pra,
  reprint,
  superscriptaddress,
  amsmath,
  amssymb,
  nofootinbib
]{revtex4-2}
\usepackage{amsfonts}
\usepackage[T1]{fontenc}
\usepackage{bm}
\usepackage{graphicx}
\usepackage{microtype}
\usepackage{xcolor}

\newcommand{\identity}{\mathbb{I}_{2}}
\newcommand{\dd}{\mathrm{d}}

\newcommand{\SLR}{\mathcal{S}_{LR}}
\newcommand{\Dph}{D_{\mathrm{ph}}}
\newcommand{\Dm}{D_{\mathrm{m}}}
\newcommand{\Rea}{\operatorname{Re}}
\newcommand{\Ima}{\operatorname{Im}}

\begin{document}

\title{Imperfect Cavities Select Better:\\ Enantioselectivity Beyond Pure Helicity}

\author{David Dams}
\affiliation{Independent Researcher}

\author{Karolina S\l{}owik}
\affiliation{Institute of Physics, Faculty of Physics, Astronomy and Informatics, Nicolaus Copernicus University in Toru\'n, Grudzi\k{a}dzka 5, 87-100 Toru\'n, Poland}
\affiliation{Institute of Advances Studies, Nicolaus Copernicus University in Toru\'n,\\ Wile\'{n}ska 4, 87-100 Toru\'n, Poland}

\date{\today}

\begin{abstract}
Single-handed light appears ideally suited to select a preferred handedness of matter. For chiral molecules in a cavity, we show that this intuition fails. Hybridization of light and matter reshapes the optical resonance through a self-energy that mixes opposite helicities. As a result, the difference in power absorbed by opposite enantiomers is generically maximized by light of mixed, rather than pure, helicity. We clarify why chiral selection admits such mixed-helicity optima in more general light--matter systems. For a concrete set of cavity modes, we find that molecular localization at a weak helicity bias considerably outperforms helically pure designs, thus identifying spatial control as a promising design direction for devices previously thought to be ``imperfect''.
\end{abstract}

\maketitle


The ability to distinguish and control the handedness of matter is a fundamental scientific challenge with far-reaching consequences across chemistry, biology, and medicine
\cite{noyori2002,blackmond2019,nguyen2006}.
At the molecular scale, discriminating between enantiomers is central to drug development and asymmetric synthesis, yet their weak chiroptical response makes optical detection difficult.
One approach exploits superchiral light to enhance chiral optical responses \cite{tang2010, tang2011, hendry2010}.
Another route is to confine light in optical cavities to strengthen its interaction with molecules and enhance chiral discrimination \cite{bougas2022}. This motivates the design \cite{solomon2019,graf2019,feis2020,both2022,mauro2023,mauro2024,duan2024,bassler2024,rebholz2026} and development \cite{semnani2020,chen2024} of helicity-preserving and single-handedness cavities \cite{voronin2022}. Achieving such helicity purity, however, is exceptionally difficult: preserving handedness over repeated reflections demands exquisite control of the mirror response and imposes stringent constraints on cavity design and fabrication.


In this work, we show that admixture of the nominally undesirable helicity does not necessarily degrade, but instead \textit{maximizes}, enantioselective performance. This is a consequence of the well-known fact that light and matter hybridize in a cavity. As a result, the same molecular couplings that generate chiral absorption also reshape the photonic resonance through a helicity-mixing self-energy, allowing helically impure light to enhance the dressed optical response in a cavity, even though it reduces the bare chiral selectivity in free space. The previously considered design space of nearly single-helicity cavities is therefore both unnecessarily restrictive and not generally optimal: helicity mixing is easier to realize than deliberately engineering cavities to support helically pure fields. These helically impure designs can, moreover, even yield stronger enantioselective responses.

To make this explicit, we formulate chiral selection as an optimization problem over a family of light--matter systems (e.g., cavities), parameterized by their helicity content
\begin{equation}
\bm{u}=\begin{pmatrix}u_+ \\
u_-\end{pmatrix},
\ \bm{u}^\dagger \bm{u}=1.
\end{equation}
We consider here a single cavity eigenmode
\begin{equation}
    \mathbf{A}(\mathbf{r}) = \bm{f}(\bm{r};\bm{u})a + \bm{f}^\star(\bm{r};\bm{u})a^\dagger
\end{equation}
of frequency $\omega_{\mathrm{ph}}$. We express the field profile $\bm{f}(\bm{r};\bm{u})$ in terms of its helical amplitudes
\begin{equation}
\bm{f}(\bm{r};\bm{u})
=
u_+ \bm{f}_+(\bm{r})
+
u_- \bm{f}_-(\bm{r}).
\label{eq:mode}
\end{equation}
We can thus think of $\bm{u}$ as identifying members of a family of cavities through appropriate boundary conditions.

Consider now a racemic mixture containing $N$ molecules of each enantiomeric species $\chi=L,R$. We model each molecule as a two-level system with transition frequency $\omega_{\mathrm{eg}}$. Within the rotating-wave approximation, the resulting family of coupled light--matter systems is described by the parameterized Tavis--Cummings Hamiltonian \cite{tavis1968,tavis1969}, as shown in Supplemental Material Sec.~\ref{sm:lrt}
\begin{equation}
H[\bm u]
=
\omega_{\mathrm{ph}}a^\dagger a
+
\omega_{\mathrm{eg}}
\sum_{i=1}^{2N}
\sigma_i^+\sigma_i^-
+
a\sum_{i=1}^{2N}
\bm u^\dagger\bm g_i\,\sigma_i^+
+
\mathrm{h.c.},
\label{eq:TC}
\end{equation}
where $a$ annihilates a photon in the cavity mode and $\sigma_i^+$ excites molecule $i$ from its ground to its excited state. The vector $\bm g_i=(g_{i,+},g_{i,-})^{\mathsf T}$ denotes the coupling of molecule $i$ to the two helicity components of the cavity mode.

As a figure of merit for chiral selection, we use the absorption difference
\begin{equation}
\SLR(\bm u,\omega)
\equiv
P_L(\bm u,\omega)-P_R(\bm u,\omega),
\end{equation}
where \(P_{\chi}(\bm u,\omega)\) denotes the power absorbed by enantiomer \(\chi\), which is directly related to its reaction rate  \cite{Stadler2018}.

Grouping all enantiomers of chirality $\chi$ in the index set $\mathcal I_\chi$, we note that their collective coupling strength to the cavity mode is given by $\sum_{i\in\mathcal I_\chi} \left|\bm u^\dagger\bm g_i\right|^2$. This can be expressed as a quadratic form in helicity space
\begin{equation}
\sum_{i\in\mathcal I_\chi}
\left|\bm u^\dagger\bm g_i\right|^2
\equiv
\bm u^\dagger M_\chi\bm u.
\label{eq:Mchi}
\end{equation}
The Hermitian $2 \times 2$ matrix $M_\chi$ admits a decomposition into Pauli matrices $\sigma^{(j)}$ with $j\in\{0,1,2,3\}$ and \mbox{$\sigma^{(0)}\equiv \identity$}
with $\sigma^{(1)}=\sigma_x$, $\sigma^{(2)}=\sigma_y$, and $\sigma^{(3)}=\sigma_z$, and appropriate real coefficients $c_j=\frac{1}{2}\mathrm{Tr}(\sigma^{(j)}M_\chi)$ quantifying the character and strength of the collective light-matter coupling in the cavity. The quadratic form is expressed in terms of the Bloch vector of the cavity mode
\begin{equation}
n_j=\bm u^\dagger\sigma^{(j)}\bm u,
\ n_1^2 + n_2^2 + n_3^2 = 1.
\label{eq:bloch}
\end{equation}
Here, $n_3=|u_+|^2-|u_-|^2$ is the helicity bias and $n_1=2\Rea(u_+^*u_-)$ describes mixing of opposite helicities.
Here, $n_3=1$ corresponds to a perfectly chiral cavity.

Exploiting time-reversal symmetry and the fact that the enantiomers are parity partners in a well-mixed racemate, we arrive at (see Appendix~\ref{sec:collective_coupling})
\begin{equation}
\bm u^\dagger M_{L/R}\bm u
=
c_0+c_1n_1\pm c_3n_3.
\label{eq:Mpauli}
\end{equation}
Having exhausted the constraints imposed by time-reversal symmetry and by the parity relation between the two enantiomeric species, we find that the cross-helicity contribution \(c_1 n_1\) remains symmetry-allowed and enters identically for both enantiomers. On the other hand, \(n_3 c_3\) is the genuinely chiral contribution to the collective coupling that distinguishes enantiomers. Without loss of generality, we can take \(c_3>0\), such that the \(L\) enantiomer couples more strongly to the positive-helicity mode.

To compute the absorption difference, we consider an external field $F$ driving the cavity mode
\begin{equation}
H_{\mathrm{drive}}
=
F(t)\left(a^\dagger+a\right).
\label{eq:drive}
\end{equation}
The bare retarded photon and molecular response kernels are
\begin{equation}
\Dph(\omega)
=
\omega-\omega_{\mathrm{ph}}+i\kappa/2,\
\Dm(\omega)
=
\omega-\omega_{\mathrm{eg}}+i\gamma/2.
\label{eq:Ddefs}
\end{equation}
Here $\kappa$ and $\gamma$ denote the photonic and molecular loss rates, respectively. A standard linear response calculation detailed in the Supplemental Material Sec.~\ref{sec:absorbed_power} yields for the cavity field
\begin{equation}
\langle a\rangle(\omega)=
\frac{F(\omega)}
{\Dph(\omega)-\Sigma(\bm u,\omega)},
\label{eq:cavityresponse}
\end{equation}
where the bare response kernel is shifted by a molecule-induced self-energy
\begin{eqnarray}\label{eq:Sigma}
\Sigma(\bm u,\omega)&=&\Dm^{-1}(\omega)\,
\bm u^\dagger(M_L+M_R)\bm u\\
&=&
2\Dm^{-1}(\omega)\left(c_0+c_1n_1\right).\nonumber
\end{eqnarray}
As the self-energy originates from the collective response of the \textit{entire} molecular ensemble, it inherits the cross-helicity coupling $c_1 n_1$, common to both enantiomers. The chiral coupling $c_3 n_3$ cancels out. Therefore, the helicity-mixing resonance shift itself is an achiral phenomenon and not restricted by the weak chiral response characteristic of realistic molecules. Since the photon field acts as an effective drive for the molecules, we finally arrive at a form combining explicitly the properties of light via $n_j$ and of its coupling to matter through $c_j$ (see Appendix~\ref{sec:absorbed_power} for the derivation)
\begin{equation}
\SLR(\bm u,\omega)
=2\mathcal K(\omega)\,
\frac{c_3n_3}
{\left|
\Dph(\omega)
-
2\Dm^{-1}(\omega)(c_0+c_1n_1)
\right|^2},
\label{eq:deltaP}
\end{equation}
where \(\mathcal K(\omega) = -\omega\,\Ima[\Dm^{-1}(\omega)]\,|F(\omega)|^2> 0 \). The numerator in Eq.~\eqref{eq:deltaP} represents the chiral bias, whose magnitude is determined by the chiral coefficient \(c_3\). Cavity dressing modifies this response through the denominator, which contains both the helicity-independent contribution \(c_0\) and the cross-helicity coupling \(c_1n_1\). While the chiral bias is thus indeed maximized at pure helicity \(|n_3| = 1\), the dressed response shifts this optimum towards mixed helicity.

Equation~\eqref{eq:deltaP} illustrates a more general principle: maximizing the electromagnetic helicity bias need not maximize the chiral selection of matter. Consider a realization of a selection functional  $\SLR^\mathrm{arb}$ in terms of an antisymmetric comparison of two responses of a time-reversal invariant system. As shown in Supplemental Material Sec.~\ref{sm:regularity}, if the two responses are exchanged by reversal of the light's handedness, the selection functional needs to be even in $n_2$ and odd in $n_3$, resulting in
\begin{equation}\label{eq:selection}
\SLR^\mathrm{arb}=n_3A(n_1,n_2^2,n_3^2).
\end{equation}
Thus, selection vanishes at $n_3=0$, but its magnitude also depends on the symmetry-allowed variation of $A$, including its dependence on the cross-helicity coordinate $n_1$.

For an unnormalized absorption difference that is quadratic in a prescribed external probe field, these same constraints imply $\SLR^\mathrm{arb}\propto n_3$, guaranteeing a pure-helicity optimum, as in the bare numerator of Eq.~\eqref{eq:deltaP}. Dynamical light--matter dressing, nonlinear response, or normalized absorption measures such as circular dichroism \cite{solomon2019} can lift this quadratic restriction and allow mixed-helicity optima. The ``superchiral'' enhancement of Tang and Cohen \cite{tang2010, tang2011} exemplifies the last mechanism.


\begin{figure*}[t!]
\centering
\includegraphics[width=\textwidth]{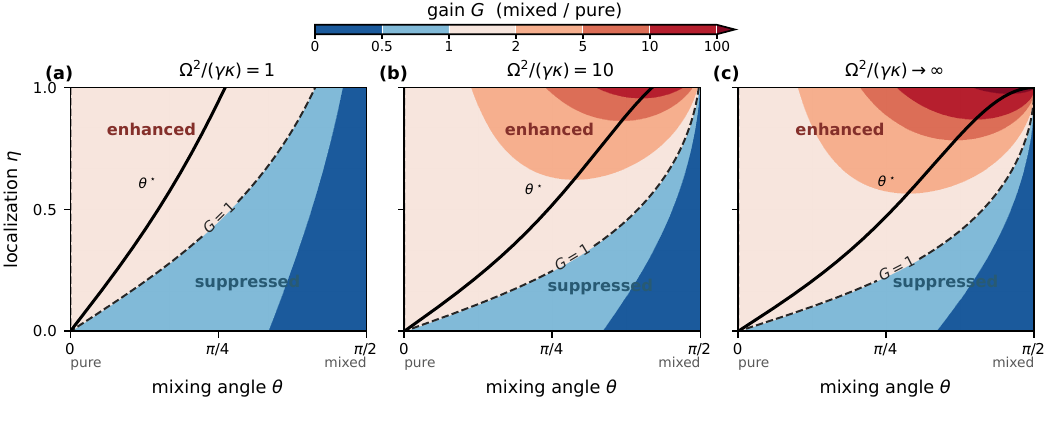}
\caption{Helicity-mixing enhancement of chiral selection.
The gain $G(\theta,\eta)=\SLR(\theta,\eta)/\SLR(0,\eta)$
relative to pure helicity at the same coupling strength is shown
as a function of mixing angle $\theta$ and localization $\eta$:
(a) $\Omega^2/(\gamma\kappa)=1$, which maximizes the resonant
selection for pure helicity;
(b) $\Omega^2/(\gamma\kappa)=10$; and
(c) the strong-coupling limit $\Omega^2/(\gamma\kappa)\to\infty$.
All panels share the same color scale, whose upper extension
denotes $G>100$.
The solid black curve marks the optimal mixing angle
$\theta^\star(\eta)$, and the dashed contour marks $G=1$.
Above this contour, helicity mixing enhances selection relative
to pure helicity. Pure helicity is optimal only at $\eta=0$;
increasing localization shifts the optimum toward mixed helicity
and increases the maximum gain.
The finite magnetic-dipole correction $B/A<1$ is retained with
$|\bm m|/(c|\bm d|)=1.4\times10^{-2}$, regularizing the
strong-coupling gain at $\eta=1$.}
\label{fig}
\end{figure*}


This effect is explicit for the absorbed power difference \(\SLR\) in Eq.~\eqref{eq:deltaP}: the dressed denominator shifts the optimum away from pure helicity and toward a mixed-helicity state. To locate this optimum, first note that our choice \(c_3>0\) selects the positive-helicity regime, \(n_3>0\). Because \(\SLR\) is maximized at \(n_2=0\), we may then write \(n_1 = -\sin(\theta), n_3= \cos(\theta)\). Normalizing the chiral selection to its value at pure positive helicity, \(\theta=0\), we obtain
\begin{equation}
G(\theta,\omega)
\equiv
\frac{\SLR(\theta,\omega)}{\SLR(0,\omega)}
=
\frac{\cos(\theta)}
{\left|1+\alpha(\omega)\sin(\theta)\right|^2},
\label{eq:gain}
\end{equation}
where \(\alpha(\omega) = \frac{2c_1\Dm^{-1}(\omega)} {\Dph(\omega)-2c_0\Dm^{-1}(\omega)}\). Linearizing this expression around \(\theta = 0\) results in
\begin{equation}
G(\theta,\omega)
=
1
-
2\Rea[\alpha(\omega)]\,\theta
+
\mathcal O(\theta^2).
\label{eq:gain-expansion}
\end{equation}
Thus, for any \(\Rea[\alpha(\omega)]\neq 0\), pure helicity is not optimal: one can always choose a nonzero \(\theta\) such that \(G(\theta,\omega)>1\). At exact light--matter resonance, \(\omega=\omega_{\mathrm{eg}}=\omega_{\mathrm{ph}}\), the mixing parameter is purely real,
\begin{equation}
\alpha
=
-\frac{8c_1}{\gamma\kappa+8c_0},
\label{eq:alpha-resonant}
\end{equation}
and Eq.~\eqref{eq:gain} can be optimized analytically. The maximum occurs at
\begin{equation}
\theta^{\star}
=
\sin^{-1}\!\left(\frac{-4\alpha}
{1+\sqrt{1+8\alpha^2}}\right).
\label{eq:n1max}
\end{equation}
Hence any nonzero cross-helicity coupling shifts the resonant optimum away from pure helicity, with larger values of \(c_1\) resulting in stronger helicity mixing. Since no fundamental principle requires \(c_1\) to be perturbatively small in realistic molecules, cross-helicity coupling can qualitatively reverse the conventional design objective.

  \begin{figure}[t]
\centering
\includegraphics[width=\columnwidth]{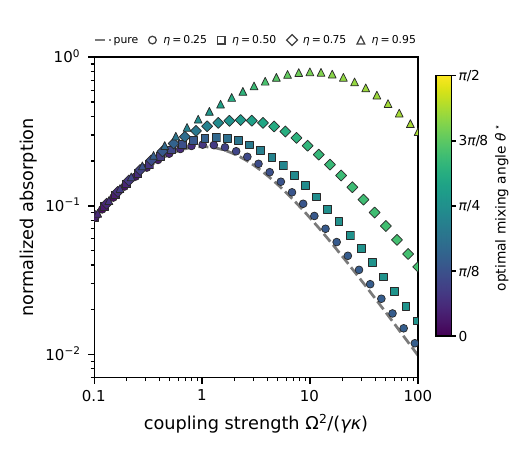}
\caption{Coupling dependence of optimized chiral absorption. The normalized absorption $\bigl[\kappa c_0/(2c_3)\bigr]\mathcal{S}_{LR}/(\omega|F|^2)$ is shown at exact light--matter resonance as a function of the dimensionless coupling strength $\Omega^2/(\gamma\kappa)$. For each localization $\eta$, the mixing angle is set to its optimum $\theta^\star(\eta,\Omega)$; marker shapes distinguish $\eta=0.25$, $0.50$, $0.75$, and $0.95$, while the marker color gives the corresponding $\theta^\star$. The dashed gray curve is the pure-helicity response at $\theta=0$,
which coincides with the result for delocalized molecules $\eta=0$.
Helicity mixing enhances the absorption above this reference for every nonzero $\eta$, with stronger localization producing both a larger enhancement and a shift of the absorption maximum toward stronger coupling.}
\label{fig:coupling-sweep}
\end{figure}

As an illustration, consider a prototypical standing-wave cavity with $\sigma=\pm$ helicity eigenmodes \cite{schafer2023,riso2023}
\begin{equation}
\bm f_\sigma(z)=f_0
\begin{pmatrix}
\cos kz\\
-\sigma\sin kz\\
0\end{pmatrix},
\ \bm\nabla\times\bm f_\sigma=\sigma k\bm f_\sigma.
\label{eq:standing}
\end{equation}
The polarization-dependent reflection boundary conditions realizing these modes are given in Supplemental Material Sec.~\ref{sm:boundaries}. Although the two helicity components are orthogonal when averaged over the full cavity, they interfere locally,
\begin{equation}
\bm f_+^*(z)\cdot\bm f_-(z)
=
|f_0|^2\cos(2kz).
\label{eq:overlap-local}
\end{equation}
A spatially confined molecular ensemble can therefore retain a finite cross-helicity overlap.

Expanding the light--matter interaction about each molecular position to electric- and magnetic-dipole order gives, as derived in Supplemental Material Sec.~\ref{sm:dipole},
\begin{equation}
g_{i,\sigma}
=
i\omega_{\mathrm{ph}}\,
\bm d_i\cdot\bm f_\sigma(\bm r_i)
+
\sigma k\,
\bm m_i\cdot\bm f_\sigma(\bm r_i),
\label{sm:eq:dipole-coupling}
\end{equation}
where $\bm d_i$ and $\bm m_i$ are the electric and magnetic transition dipoles. For opposite enantiomers the transition moments may be chosen such that $\bm d_L=\bm d_R$ and $\bm m_L=-\bm m_R$. Hence the achiral magnitudes $|\bm d|$ and $|\bm m|$ are identical for both species.

Let $\rho(\bm r)$ denote the unit-normalized spatial distribution, taken to be the same for both enantiomers, and assume random molecular orientations. Orientational averaging then gives for the collective couplings entering the self-energy (see Appendix~\ref{sm:dipole})
\begin{align}
c_0
&=
\frac{N}{3}
\left(
\omega_{\mathrm{ph}}^2|\bm d|^2
+
k^2|\bm m|^2
\right)|f_0|^2
\equiv
\frac{N}{3}A|f_0|^2,
\\
c_1
&=
\frac{N}{3}
\left(
\omega_{\mathrm{ph}}^2|\bm d|^2
-
k^2|\bm m|^2
\right)|f_0|^2\,\eta
\equiv
\frac{N}{3}B|f_0|^2\,\eta.
\end{align}
Here,
\begin{equation}
\eta
\equiv
\int\dd\bm r\,\rho(\bm r)\cos(2kz).
\label{eq:eta}
\end{equation}
quantifies the cross-helicity interference sampled by the molecules.
The relative strength of the cross-helicity and helicity-independent collective couplings is therefore
\begin{equation}
\frac{c_1}{c_0}
=
\frac{B}{A}\,\eta.
\label{eq:c1c0}
\end{equation}
For ordinary molecular transitions with $k|\bm m|\ll\omega_{\mathrm{ph}}|\bm d|$, one has $B/A\simeq1$, so helicity mixing is controlled primarily by the spatial factor $\eta$. Spatially extended ensembles average the opposite-helicity interference away, whereas localized ensembles retain it.
At light--matter resonance, the corresponding mixing parameter becomes
\begin{equation}
\alpha
=
-\frac{B}{A}
\frac{\eta}{1+\gamma\kappa/\Omega^2},
\qquad
\Omega^2
=
\frac{8N}{3}A|f_0|^2.
\label{eq:q}
\end{equation}
For ideal perfectly chiral molecules, both $c_1$ and $B$ would vanish, maximizing the differential absorption at pure helicity. For realistic molecules,
both the spatially sampled interference and the molecular ratio $B/A$ control the self-energy induced by helicity mixing, while $\Omega$ sets the collective light--matter coupling scale and is analogous to the Rabi splitting. For a pure-helicity mode, the resonant chiral selection is maximized at the critical coupling $\Omega^2/(\gamma\kappa)=1$. Increasing the coupling beyond this point does not uniformly enhance \(\SLR\): in the strong-coupling regime, the resonance splits into two polaritonic branches, shifting spectral weight away from the bare molecular excitation frequency. We discuss the dependence on \(\Omega\) and the strong-coupling limit in Supplemental Material Sec.~\ref{sm:strong}.

Against this pure-helicity optimum, Fig.~\ref{fig} shows how spatially sampled cross-helicity overlap provides an additional route to enhancement. Rather than being merely an imperfection to suppress, $\eta$ acts as a control parameter through which the molecular self-energy can compensate for reduced helicity bias and enhance the dressed chiral response beyond that of a helically pure mode at the same coupling strength.
Spatial control of the molecular ensemble therefore both relaxes the demanding requirement of fabricating nearly single-helicity cavities and opens a regime in which such cavities would not be optimal even if they were available.

Because an enhanced gain does not by itself imply a large absolute response, Fig.~\ref{fig:coupling-sweep} also shows the optimized absorption normalized to the incident power over a broad coupling range. The mixed-helicity response remains above the pure-helicity reference for every nonzero localization considered, while increasing $\eta$ produces a larger enhancement and shifts the absorption maximum both toward stronger coupling and helicity mixing. Thus, the advantage of helicity mixing is not restricted to the formal strong-coupling limit but persists throughout a broad finite-coupling regime.

For example, a homogeneous molecular slab of thickness $D$ centered at $z_0$ samples
\[
\eta
=
\frac{\sin(kD)}{kD}\cos(2kz_0),
\]
so both localization and positioning provide direct experimental control over the helicity-mixing self-energy.

In conclusion, we have shown that light--matter hybridization changes the design principles for enantioselective cavities. Maximizing helicity purity optimizes chiral selection in the linear, undressed limit, but the interacting system admits an additional control mechanism: the same molecular couplings that produce differential absorption also reshape the cavity resonance. This dressing generically shifts the optimal chiral response toward mixed helicity, making helicity purity an unnecessarily restrictive design objective. For the standing-wave modes considered here, molecular localization retains interference between opposite helicities that would otherwise average to zero, allowing the sample position and thickness to control the dressed response. The resulting enhancement persists at finite coupling and enables mixed-helicity modes to outperform pure-helicity modes at the same collective coupling strength. Optimizing the joint light--matter response therefore opens a broader design space that can relax the demanding requirements of helically pure cavities: designs previously regarded as imperfect may offer superior enantioselective performance.

\bibliography{cavities}
\appendix
\setcounter{section}{0}
\renewcommand{\thesection}{\Alph{section}}
\renewcommand{\theequation}{\thesection\arabic{equation}}
\setcounter{equation}{0}
\section{Derivation of the Hamiltonian}\label{sm:lrt}
The cavity vector potential $\mathbf{A}(\mathbf{r})$ and the associated electric $\mathbf{E}(\mathbf{r})$ and magnetic $\mathbf{B}(\mathbf{r})$ fields can be expressed as
\begin{align}
\bm A(\bm r)&=\bm f(\bm r;\bm u)a+\bm f^*(\bm r;\bm u)a^\dagger,\\
\bm E(\bm r)&=i\omega_{\mathrm{ph}}\left[\bm f(\bm r;\bm u)a-\bm f^*(\bm r;\bm u)a^\dagger\right],\\
\bm B(\bm r)&=\nabla\times\bm A(\bm r),
\end{align}
where the mode annihilation operator fulfills the standard commutation relation $[a,a^\dagger]=1$.

The cavity is filled with a racemic mixture of molecules numbered by the index $i$ and described by the collective Hamiltonian $H_\mathrm{mol}$. The full Hamiltonian of the system includes the interaction term, which can be expressed through the polarization $\mathbf{P}_i(\mathbf{r})$ and magnetization $\mathbf{M}_i(\mathbf{r})$, defined for each molecule \cite{steck}
\begin{equation}
H=H_{\mathrm{mol}}+\omega_{\mathrm{ph}}a^\dagger a
-\sum_i\!\int\!\dd\bm r\,\bm E\cdot\bm P_i
-\sum_i\!\int\!\dd\bm r\,\bm B\cdot\bm M_i .
\end{equation}
Projecting onto the molecular ground and excited states and applying the rotating wave approximation (RWA) yields helicity-resolved light-matter coupling constants
\begin{equation}
\begin{aligned}
g_{i,\sigma}
={}&
-i\omega_{\mathrm{ph}}
\langle e_i| \int\dd\bm r\,
f_\sigma(\bm r) \cdot \bm P_i(\bm r)|g_i\rangle
\\
&-
\langle e_i|\int\dd\bm r\,
\bigl[\nabla\times\bm f_\sigma(\bm r)\bigr]\cdot
\bm M_i(\bm r)|g_i\rangle.
\end{aligned}
\label{sm:coupling}
\end{equation}
Then, the Hamiltonian becomes
\begin{equation}
H[\bm u]
=
\omega_{\mathrm{ph}}a^\dagger a
+
\omega_{\mathrm{eg}}
\sum_{i=1}^{2N}
\sigma_i^+\sigma_i^-
+
a\sum_{i=1}^{2N}
\bm u^\dagger\bm g_i\,\sigma_i^+
+
\mathrm{h.c.},
\label{sm:TC}
\end{equation}
where
\(\sigma_i^+ = |e_i\rangle\langle g_i|\)
and
\(\sigma_i^- = |g_i\rangle\langle e_i|\)
are the molecular raising and lowering operators, respectively. This is expression (\ref{eq:TC}) for \(H\) given in the main text.

\section{Enantiomer-Resolved Collective Coupling}
\label{sec:collective_coupling}

The collective coupling strength of enantiomeric species $\chi$ can be
written as
\begin{equation}\label{eq:Mpauli2}
\begin{aligned}
\sum_{i\in\mathcal I_\chi}
\left|\bm u^\dagger\bm g_i\right|^2
&=\bm u^\dagger M_\chi\bm u\\
&=c_0^\chi+c_1^\chi n_1+c_2^\chi n_2+c_3^\chi n_3,
\end{aligned}
\end{equation}
where the real coefficients $c_j^\chi$ explicitly depend on the
enantiomeric species, and the set $\mathcal I_\chi$ includes molecules of a given enantiomer.

We consider a well-mixed racemate with equal populations and identical spatial and orientational distributions of both enantiomers.
Within the mode family considered, complete reversal of the field's handedness is represented by exchanging its helicity amplitudes
\begin{equation}
X\bm u=\sigma_x\bm u,
\qquad
X:(n_1,n_2,n_3)\mapsto(n_1,-n_2,-n_3).
\label{eq:helicity_exchange}
\end{equation}
Reversing the light's handedness interchanges the roles of the two enantiomers. Since both species sample the field identically, their
collective coupling strengths are exchanged
\begin{equation}
(X\bm u)^\dagger M_L(X\bm u)
=\bm u^\dagger M_R\bm u.
\label{eq:coupling_exchange}
\end{equation}
Comparing coefficients in Eq.~\eqref{eq:Mpauli2} therefore gives
\begin{equation}
\begin{aligned}
c_0^L&=c_0^R\equiv c_0,
&\qquad c_1^L&=c_1^R\equiv c_1,\\
c_2^L&=-c_2^R\equiv c_2,
&\qquad c_3^L&=-c_3^R\equiv c_3.
\end{aligned}
\label{eq:coeffs_light}
\end{equation}

In the chosen mode basis, time reversal acts as complex conjugation:
\begin{equation}
\mathcal T\bm u=\bm u^*,
\qquad
\mathcal T:(n_1,n_2,n_3)\mapsto(n_1,-n_2,n_3).
\label{eq:coeffs_time}
\end{equation}
Time-reversal invariance implies that the collective coupling strengths
are unchanged by this transformation
\begin{equation}
(\mathcal T\bm u)^\dagger M_\chi(\mathcal T\bm u)
=\bm u^\dagger M_\chi\bm u.
\label{eq:trs}
\end{equation}
Since time reversal changes only the sign of $n_2$,
Eqs.~\eqref{eq:coeffs_time} and \eqref{eq:trs} require $c_2^\chi=0$.
Combining this result with Eq.~\eqref{eq:coeffs_light}, we obtain
\begin{equation}
\bm u^\dagger M_{L/R}\bm u
=c_0+c_1n_1\pm c_3n_3,
\label{eq:Mpauli_reduced}
\end{equation}
as stated in Eq.~\eqref{eq:Mpauli} of the main text.

\section{Absorbed power}\label{sec:absorbed_power}
Let us collect the photon and the $2N$ molecular lowering operators into $\Psi=(a,\bm\sigma^-)^{T}$. The retarded inverse Green function is given by
\begin{equation}
G^{-1}_{R,\mu\nu}(\omega)
=\langle G|\Psi_\mu\,(\omega+i0^+-H)\,\Psi_\nu^\dagger|G\rangle,
\end{equation}
where \(|G\rangle\) is the zero-excitation ground state
\begin{equation}
|G\rangle=|0\rangle_{\mathrm{ph}}\otimes\bigotimes_{i=1}^{2N}|g_i\rangle.
\end{equation}
Since the state \(\Psi_\nu^\dagger|G\rangle\) lies in the singly excited manifold, only the single-excitation sector of the Hamiltonian contributes and we have for the equation of motion with the external drive \(F\)
\begin{widetext}
\begin{equation}
  G^{-1}_{R,\mu\nu}(\omega)
  \begin{pmatrix}
    \langle a\rangle\\
    \langle\bm\sigma^-\rangle
  \end{pmatrix}
  \nonumber\\
  =  \begin{pmatrix}
    \Dph(\omega) & -\bm V^\dagger\\
    -\bm V & \Dm(\omega)\,\mathbb I_{2N}
  \end{pmatrix}
  \begin{pmatrix}
    \langle a\rangle\\
    \langle\bm\sigma^-\rangle
  \end{pmatrix}
  =  \begin{pmatrix}
    F(\omega)\\
    \bm 0
  \end{pmatrix},
\end{equation}
\end{widetext}
with $\bm V$ being a vector collecting $V_i=\bm u^\dagger\bm g_i$.
Eliminating the molecular amplitudes gives
\begin{equation}
\langle a\rangle=\frac{F}{\Dph-\Dm^{-1}\bm V^\dagger\bm V},
\qquad
\langle\sigma_i^-\rangle=\Dm^{-1}V_i\langle a\rangle,
\end{equation}
where
\begin{equation}
\bm V^\dagger\bm V = \bm u^\dagger(M_L+M_R)\bm u .
\end{equation}
This gives the expression for the self energy \(\Sigma(\bm u, \omega)\) defined in the main text. The power absorbed by molecule \(i\) is the out-of-phase response to the effective drive $V_i\langle a\rangle$ \cite{Giuliani_Vignale_2005}, such that we have
\begin{align}
P_\chi
&=-\omega\sum_{i\in\mathcal I_\chi}
\Ima\!\left[(V_i\langle a\rangle)^*\langle\sigma_i^-\rangle\right]\\
&=-\omega\,\Ima(\Dm^{-1})\,|F|^2\,
\frac{\bm u^\dagger M_\chi\bm u}{|\Dph-\Sigma(\bm u, \omega)|^2},
\end{align}
Evaluating $P_L - P_R$ yields the expression (\ref{eq:deltaP}) in the main text.

\section{Selection Functional}
\label{sm:regularity}

Consider two responses $a(\bm n)$ and $b(\bm n)$ that are exchanged
by reversal of the light's handedness:
\begin{equation}
\begin{aligned}
    a(X\bm n)&=b(\bm n), \\
    X:(n_1,n_2,n_3)&\mapsto(n_1,-n_2,-n_3).
\end{aligned}
\end{equation}
Since $X^2=1$, this also implies $b(X\bm n)=a(\bm n)$.
In the cavity setting considered here, these are the powers absorbed
by the two enantiomeric species in a well-mixed racemate.
For an isolated molecule, they can instead denote its responses to
two fields related by handedness reversal.

Let $\SLR^\mathrm{arb}$ be a realization of a selection functional in terms of an antisymmetric comparison $\Phi$ of the two responses,
\begin{equation}
\Phi(a,b)=-\Phi(b,a),
\qquad
\SLR(\bm n)=\Phi\bigl(a(\bm n),b(\bm n)\bigr).
\end{equation}
We additionally assume that both responses are invariant under
time reversal, represented in the chosen mode basis by
\begin{equation}
\mathcal T:(n_1,n_2,n_3)\mapsto(n_1,-n_2,n_3).
\end{equation}
It follows that
\begin{equation}
\SLR(n_1,n_2,n_3)=\SLR(n_1,-n_2,n_3),
\end{equation}
so $\SLR$ is even in $n_2$.
Handedness reversal exchanges the two responses and therefore gives
\begin{equation}
\begin{aligned}
\SLR(X\bm n)
&=\Phi\bigl(b(\bm n),a(\bm n)\bigr)\\
&=-\SLR(\bm n).
\end{aligned}
\end{equation}
Combining these relations yields
\begin{equation}
\SLR(n_1,n_2,-n_3)=-\SLR(n_1,n_2,n_3),
\end{equation}
so $\SLR$ is odd in $n_3$.
These symmetry constraints imply
\begin{equation}
\SLR(\bm n)=n_3\,A(n_1,n_2^2,n_3^2),
\label{eq:restricted}
\end{equation}
as stated in the main text.

For a prescribed external field
$\bm F=u_+\bm F_++u_-\bm F_-$, the unnormalized absorption
difference in linear response is a quadratic form,
\begin{equation}
\SLR
=\bm F^\dagger M_{LR}\bm F
=\bm u^\dagger
m_{LR}[\bm F_+,\bm F_-]\bm u,
\end{equation}
Compatibility with Eq.~\eqref{eq:restricted} then requires
$\SLR\propto n_3$.
For nonzero proportionality coefficient, the magnitude of the
selection is therefore maximized at pure helicity.

In the cavity setting, light--matter coupling makes the driven
field amplitude depend on the mode composition through the
self-energy. The resulting absorption difference need no longer
be quadratic in $\bm u$ and can acquire the symmetry-allowed
dependence on $n_1$.
Such dependence can also arise through normalization: the
dissymmetry factor used to characterize superchiral light is
a ratio of absorption responses, even when each response
separately is quadratic in the field.

\section{Boundary Conditions for the Standing-Wave Cavity}
\label{sm:boundaries}

We now show explicitly that the standing-wave modes considered in the main text can be realized by suitable polarization-dependent boundary conditions. We write the cavity mode as
\begin{equation}
\bm f(\bm r;\bm u)
=
u_+\bm f_+(\bm r)
+
u_-\bm f_-(\bm r),
\label{sm:eq:mode-superposition}
\end{equation}
where \(\bm u=(u_+,u_-)^{\mathsf T}\) specifies the helicity composition.

Helicity is the projection of spin onto the direction of propagation. Accordingly, a standing wave of helicity \(\sigma=\pm\) can be decomposed into two counterpropagating waves with opposite circular polarizations. Defining
\begin{equation}
\bm e_\sigma
=
\frac{1}{\sqrt{2}}
\begin{pmatrix}
1\\
i\sigma
\end{pmatrix},
\end{equation}
we may write
\begin{equation}
\bm f_\sigma(\bm r)
=
\frac{f_0}{\sqrt{2}}
\left(
\bm e_\sigma e^{ikz}
+
\bm e_{-\sigma}e^{-ikz}
\right).
\label{sm:eq:helicity-standing}
\end{equation}
This is equivalent to the standing-wave form used in the main text,
\begin{equation}\label{eq:cavity_mode}
\bm f_\sigma(z)
=
f_0
\begin{pmatrix}
\cos kz\\
-\sigma\sin kz
\end{pmatrix}.
\end{equation}

Equation~\eqref{sm:eq:mode-superposition} can therefore be written as a superposition of forward- and backward-propagating waves,
\begin{equation}
\bm f(\bm r;\bm u)
=
\frac{f_0}{\sqrt{2}}
\left[
\bm p_+(\bm u)e^{ikz}
+
\bm p_-(\bm u)e^{-ikz}
\right],
\label{sm:eq:traveling-decomposition}
\end{equation}
with polarization vectors
\begin{equation}
\begin{aligned}
\bm p_+(\bm u)
&=
u_+\bm e_+
+
u_-\bm e_-,
\\
\bm p_-(\bm u)
&=
u_-\bm e_+
+
u_+\bm e_-.
\end{aligned}
\label{sm:eq:p-polarizations}
\end{equation}
Although each vector is normalized for \(\bm u^\dagger\bm u=1\), the forward- and backward-propagating polarizations are in general not orthogonal. In fact,
\begin{equation}
\bm p_+^\dagger\bm p_-
=
2\operatorname{Re}(u_+^*u_-)
=
n_1.
\end{equation}

Consider a cavity formed by two polarization-dependent mirrors, represented by reflection matrices \(R_{\rightarrow}(\bm u)\) and \(R_{\leftarrow}(\bm u)\). For a given design vector \(\bm u\), we require the mirrors to map the polarization of each incident traveling wave onto that of its counterpropagating partner,
\begin{equation}
\begin{aligned}
R_{\rightarrow}(\bm u)\bm p_-(\bm u)
&=
e^{i\phi_{\rightarrow}}\bm p_+(\bm u),
\\
R_{\leftarrow}(\bm u)\bm p_+(\bm u)
&=
e^{i\phi_{\leftarrow}}\bm p_-(\bm u).
\end{aligned}
\label{sm:eq:reflection-desired}
\end{equation}

Let \(\bm p_\pm^\perp\) denote normalized polarization vectors orthogonal to \(\bm p_\pm\). A convenient unitary realization of the reflection matrices is then
\begin{equation}
\begin{aligned}
R_{\rightarrow}(\bm u)
={}&
e^{i\phi_{\rightarrow}}
\bm p_+(\bm u)\bm p_-^\dagger(\bm u)
+
e^{i\psi_{\rightarrow}}
\bm p_+^\perp(\bm u)
\bm p_-^{\perp\dagger}(\bm u),
\\
R_{\leftarrow}(\bm u)
={}&
e^{i\phi_{\leftarrow}}
\bm p_-(\bm u)\bm p_+^\dagger(\bm u)
+
e^{i\psi_{\leftarrow}}
\bm p_-^\perp(\bm u)
\bm p_+^{\perp\dagger}(\bm u).
\end{aligned}
\label{sm:eq:reflection-matrices}
\end{equation}
The phases \(\phi_{\rightarrow},\phi_{\leftarrow}\) determine the resonance of the desired polarization channel, whereas \(\psi_{\rightarrow}, \psi_{\leftarrow}\) determine that of the orthogonal channel.

For a cavity of length \(L\), a forward-propagating wave undergoes the round trip
\begin{equation}
\bm p_+
\longmapsto
e^{2ikL}R_{\rightarrow}R_{\leftarrow}\bm p_+
=
e^{i(2kL+\phi_{\rightarrow}+\phi_{\leftarrow})}\bm p_+.
\end{equation}
The corresponding condition for a backward-propagating wave follows from the reversed ordering \(R_{\leftarrow}R_{\rightarrow}\) and gives the same accumulated phase. The resonance condition for the desired cavity mode is therefore
\begin{equation}
2k_mL+\phi_{\rightarrow}+\phi_{\leftarrow}=2\pi m,
\end{equation}
or
\begin{equation}
k_m
=
\frac{2\pi m-\phi_{\rightarrow}-\phi_{\leftarrow}}{2L}.
\label{sm:eq:resonance-k}
\end{equation}
Accordingly,
\begin{equation}
\omega_{\mathrm{ph}}
=
\frac{c}{2L}
\left(
2\pi m-\phi_{\rightarrow}-\phi_{\leftarrow}
\right).
\label{sm:eq:resonance-omega}
\end{equation}
The orthogonal polarization channel instead satisfies
\begin{equation}
2kL+\psi_{\rightarrow}+\psi_{\leftarrow}=2\pi m'.
\end{equation}
Thus, choosing
\begin{equation}
\phi_{\rightarrow}+\phi_{\leftarrow}
\neq
\psi_{\rightarrow}+\psi_{\leftarrow}
\pmod{2\pi}
\end{equation}
separates the two polarization channels spectrally, allowing the desired mode to remain resonant at \(\omega_{\mathrm{ph}}\) while the orthogonal mode is shifted away from resonance.

For completeness, the reflection matrices can be written explicitly. Define
\begin{equation}
a=u_++u_-,
\qquad
b=u_+-u_-.
\end{equation}
The traveling-wave polarization vectors are
\begin{equation}
\begin{aligned}
\bm p_+(\bm u)
&=
\frac{1}{\sqrt{2}}
\begin{pmatrix}
a\\
ib
\end{pmatrix},
&
\bm p_-(\bm u)
&=
\frac{1}{\sqrt{2}}
\begin{pmatrix}
a\\
-ib
\end{pmatrix},
\\[1ex]
\bm p_+^\perp(\bm u)
&=
\frac{1}{\sqrt{2}}
\begin{pmatrix}
b^*\\
-ia^*
\end{pmatrix},
&
\bm p_-^\perp(\bm u)
&=
\frac{1}{\sqrt{2}}
\begin{pmatrix}
b^*\\
ia^*
\end{pmatrix}.
\end{aligned}
\label{sm:eq:p-explicit}
\end{equation}

Substituting these expressions into Eq.~\eqref{sm:eq:reflection-matrices} gives
\begin{equation}
R_{\rightarrow}(\bm u)
=
\frac{1}{2}
\begin{pmatrix}
e^{i\phi_{\rightarrow}}|a|^2+e^{i\psi_{\rightarrow}}|b|^2
&
i\,ab^*
\left(e^{i\phi_{\rightarrow}}-e^{i\psi_{\rightarrow}}\right)
\\[1ex]
i\,a^*b
\left(e^{i\phi_{\rightarrow}}-e^{i\psi_{\rightarrow}}\right)
&
-e^{i\phi_{\rightarrow}}|b|^2-e^{i\psi_{\rightarrow}}|a|^2
\end{pmatrix},
\label{sm:eq:RL-explicit}
\end{equation}
and
\begin{equation}
R_{\leftarrow}(\bm u)
=
\frac{1}{2}
\begin{pmatrix}
e^{i\phi_{\leftarrow}}|a|^2+e^{i\psi_{\leftarrow}}|b|^2
&
-i\,ab^*
\left(e^{i\phi_{\leftarrow}}-e^{i\psi_{\leftarrow}}\right)
\\[1ex]
-i\,a^*b
\left(e^{i\phi_{\leftarrow}}-e^{i\psi_{\leftarrow}}\right)
&
-e^{i\phi_{\leftarrow}}|b|^2-e^{i\psi_{\leftarrow}}|a|^2
\end{pmatrix}.
\label{sm:eq:RR-explicit}
\end{equation}
Finally, using
\begin{equation}
ab^*
=
|u_+|^2-|u_-|^2
+
2i\,\operatorname{Im}(u_+^*u_-)
=
n_3+in_2,
\end{equation}
we see that \(n_2=0\) implies \(ab^*=a^*b\). Consequently, in the fixed transverse polarization basis used above,
\begin{equation}
R_{\rightarrow}^{\mathsf T}=R_{\rightarrow},
\qquad
R_{\leftarrow}^{\mathsf T}=R_{\leftarrow},
\end{equation}
which is the symmetry condition for reciprocal Jones reflection matrices. Thus the \(n_2=0\) cavity family relevant to the optimum discussed in the main text can be realized using reciprocal polarization-dependent mirrors.

\section{Electric- and magnetic-dipole coupling}
\label{sm:dipole}
We now derive the coefficients \(c_0,c_1,c_3\) entering Eq.~\eqref{eq:Mpauli} from the electric- and magnetic-dipole interaction for a generic homogeneous cavity characterized by a constant real refractive index $n$, before specializing to the standing-wave cavity considered in the main text. We consider the helicity-resolved coupling in Eq.~\eqref{sm:coupling}. Expanding the molecular polarization and magnetization about the molecular position \(\bm r_i\), the coupling \(g_{i, \sigma}\) of molecule \(i\) to helicity \(\sigma=\pm\) takes, to dipolar order, the form
\begin{equation}
g_{i,\sigma}
=
i\omega_{\mathrm{ph}}\,
\bm d_i\cdot\bm f_\sigma(\bm r_i)
+
\sigma k\,
\bm m_i\cdot\bm f_\sigma(\bm r_i),
\end{equation}
where we have used
\begin{equation}
\nabla\times\bm f_\sigma
=
\sigma k\,\bm f_\sigma,
\qquad
k=\frac{n\omega_{\mathrm{ph}}}{c},
\end{equation}
which holds for helicity eigenmodes in a source-free homogeneous region. The electric and magnetic transition dipoles are
\begin{equation}
\bm d_i
=
\langle e_i|\bm P (\bm r_i) |g_i\rangle,
\qquad
\bm m_i
=
\langle e_i|\bm M (\bm r_i) |g_i\rangle.
\end{equation}

For two appropriately aligned enantiomers of opposite handedness, the molecular index $i$ can be dropped and the transition moments may be chosen such that
\begin{equation}
\bm d_L=\bm d_R,
\qquad
\bm m_L=-\bm m_R.
\end{equation}

Let the unit-normalized molecular density \(\rho(\bm r)\) be the same for both enantiomers, and assume that the molecules are randomly oriented. We may then introduce representative transition dipoles \(\bm d_\chi\) and \(\bm m_\chi\) for enantiomer \(\chi=L,R\), and define the position-dependent helicity coupling
\begin{equation}
g_{\chi\sigma}(\bm r)
=
i\omega_{\mathrm{ph}}\,
\bm d_\chi\cdot\bm f_\sigma(\bm r)
+
\sigma k\,
\bm m_\chi\cdot\bm f_\sigma(\bm r).
\end{equation}
The phases of the molecular states may be chosen such that \(\bm d_\chi\) is real and \(\bm m_\chi\) is purely imaginary.

For \(N\) randomly oriented molecules of each enantiomer, distributed according to \(\rho(\bm r)\), the collective coupling matrix introduced in Eq.~\eqref{main-eq:Mchi} becomes
\begin{equation}
M_\chi
=
N\int\dd\bm r\,\rho(\bm r)\,
\left\langle
\bm g_\chi(\bm r)
\bm g_\chi^\dagger(\bm r)
\right\rangle_{\mathrm{or}},
\label{sm:eq:Mchi-dipole}
\end{equation}
where
\begin{equation}
\bm g_\chi(\bm r)
=
\begin{pmatrix}
g_{\chi+}(\bm r)\\
g_{\chi-}(\bm r)
\end{pmatrix},
\end{equation}
and \(\langle\cdots\rangle_{\mathrm{or}}\) denotes the orientational average.

For two molecular vectors \(\bm v\) and \(\bm w\) undergoing the same random rotation, isotropy gives
\begin{equation}
\left\langle
(\bm v\cdot\bm f_\sigma)
(\bm w\cdot\bm f_{\sigma'})^*
\right\rangle_{\mathrm{or}}
=
\frac{1}{3}
(\bm v\cdot\bm w^*)\,
(\bm f_\sigma\cdot\bm f_{\sigma'}^*).
\end{equation}
It is therefore convenient to introduce the molecularly sampled mode overlaps
\begin{equation}
S_{\sigma\sigma'}
=
\int\dd\bm r\,\rho(\bm r)\,
\bm f_\sigma^*(\bm r)\cdot
\bm f_{\sigma'}(\bm r).
\label{sm:eq:S}
\end{equation}

With the enantiomer-independent magnitudes

\begin{equation}
|\bm d|^2
\equiv
|\bm d_L|^2
=
|\bm d_R|^2,
\qquad
|\bm m|^2
\equiv
|\bm m_L|^2
=
|\bm m_R|^2
\end{equation}
we can additionally define the molecular scales
\begin{equation}
A
=
\omega_{\mathrm{ph}}^2|\bm d|^2
+
k^2|\bm m|^2,
\qquad
B
=
\omega_{\mathrm{ph}}^2|\bm d|^2
-
k^2|\bm m|^2.
\label{sm:eq:AB}
\end{equation}

Molecular chirality enters through the rotatory strength
\begin{equation}
R_\chi
=
\Ima\!\left(
\bm d_\chi^*\cdot\bm m_\chi
\right),
\qquad
R_L=-R_R\equiv R.
\label{sm:eq:rotatory}
\end{equation}
The microscopic coupling matrices therefore take the form
\begin{align}
M_L
&=
\frac{N}{3}
\begin{pmatrix}
(A+2\omega_{\mathrm{ph}}kR)S_{++} & B S_{+-}\\
B S_{-+} & (A-2\omega_{\mathrm{ph}}kR)S_{--}
\end{pmatrix},
\\
M_R
&=
\frac{N}{3}
\begin{pmatrix}
(A-2\omega_{\mathrm{ph}}kR)S_{++} & B S_{+-}\\
B S_{-+} & (A+2\omega_{\mathrm{ph}}kR)S_{--}
\end{pmatrix}.
\end{align}

  \begin{figure}[t!]
\centering
\includegraphics[width=\columnwidth]{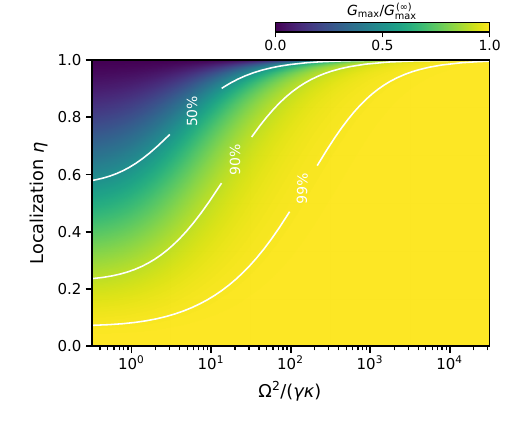}
\caption{Maximum gain \(G(\theta^{\star}, \eta)\) normalized to its strong-coupling value (at $\Omega\rightarrow \infty$), as a function of the coupling strength \(\frac{\Omega^2}{\gamma\kappa}\) and localization parameter \(\eta\). The contour lines indicate where 50, 90 and 99 percent of this maximum value are reached. For \(\eta \sim 0.1\), saturation sets in for coupling strengths below 1, while  \(\eta > 0.5\) saturates at coupling strengths above 100.}
\label{fig:gainlimit}
\end{figure}

We now compare these expressions with the symmetry-constrained form
derived in the main text,
\begin{equation}
\bm u^\dagger M_{L/R}\bm u
=
c_0+c_1n_1\pm c_3n_3 ,
\end{equation}
from which it follows that
\begin{equation}
M_{L/R}
=
\begin{pmatrix}
c_0\pm c_3 & c_1\\
c_1 & c_0\mp c_3
\end{pmatrix},
\end{equation}
From this, we immediately see that \(S_{++}=S_{--}\equiv S\) as well as \(S_{+-} = S_{-+}\) is real. We finally arrive at
\begin{align}
c_0 &= \frac{N}{3}A\,S, \label{eq:cs:first}\\
c_1 &= \frac{N}{3}B\,S_{+-},\\
c_3 &= \frac{2N}{3}\omega_{\mathrm{ph}}kR\,S. \label{eq:cs:last}
\end{align}

The discussion above applies to a generic source-free homogeneous region. We now specialize to the standing-wave modes of Eq.~\eqref{eq:cavity_mode}.
Pointwise,
\begin{equation}
\begin{aligned}
|\bm f_\sigma(z)|^2&=|f_0|^2,
\bm f_+^*(z)\cdot\bm f_-(z) \\
&=|f_0|^2\cos(2kz).
\end{aligned}
\end{equation}
It follows from (\ref{sm:eq:S}) that
\begin{equation}
S=|f_0|^2, \
S_{+-}=|f_0|^2\eta,
\ \eta=\int\dd\bm r\,\rho(\bm r)\cos(2kz).
\end{equation}
Equations~\eqref{eq:cs:first}--\eqref{eq:cs:last} therefore reduce to
\begin{align}
c_0
&=
\frac{N}{3}
\left(
\omega_{\mathrm{ph}}^2|\bm d|^2
+
k^2|\bm m|^2
\right)|f_0|^2,
\\
c_1
&=
\frac{N}{3}
\left(
\omega_{\mathrm{ph}}^2|\bm d|^2
-
k^2|\bm m|^2
\right)|f_0|^2\eta,
\end{align}
as quoted in the main text. For ordinary molecular transitions $k|\bm m|\ll\omega_{\mathrm{ph}}|\bm d|$,
and therefore $\frac{B}{A}\simeq1,$
such that the relative magnitude of the helicity-independent collective coupling \(c_0\) and the cross-helicity collective coupling \(c_1\) depends on the cross-helicity interference \(\eta\) as
$\frac{c_1}{c_0} \simeq \eta,$
with \(\frac{B}{A}\) acting as a physical regularizer, preventing the divergence of the expressions for the gain, discussed in the following section, as well as in the main text.

\section{Strong coupling limit}
\label{sm:strong}

The general gain in Eq.~\eqref{eq:gain} in the main text over pure helicity at the same coupling strength is

\begin{equation}
G(\theta,\eta)
\simeq
\frac{\cos\theta}
{\left[
1-\dfrac{B}{A}
\dfrac{\eta}{1+\gamma\kappa/\Omega^2}
\sin\theta
\right]^2}.
\label{eq:cfull}
\end{equation}

The effective mixing parameter $c_1$ is therefore controlled by both the localization $\eta$ and the dimensionless coupling $\Omega^2/(\gamma\kappa)$. Increasing the coupling strengthens this shift and moves the optimum toward larger mixing angles $\theta$. In the collective strong-coupling limit, $\Omega^2\gg\gamma\kappa$, Eq.~\eqref{eq:cfull} reduces to
\begin{equation}
G(\theta,\eta)
\simeq
\frac{\cos\theta}
{\left[
1-\dfrac{B}{A}\eta\sin\theta
\right]^2}.
\label{eq:Climit}
\end{equation}
We numerically demonstrate the onset of this strong-coupling limit in Fig.~\ref{fig:gainlimit}.
\end{document}